\documentclass{article}

\usepackage{amsmath}
\usepackage{amsfonts}
\usepackage{amssymb}
\usepackage{graphicx}
\usepackage{animate}
\usepackage{amsthm}
\usepackage{multirow}
\usepackage[round]{natbib}
\usepackage{lscape}
\usepackage{rotating}
\usepackage{animate}
\usepackage{setspace}

\begin{document}

\title{Estimation and approximation in multidimensional dynamics}
\author{Gianluca Frasso, Jonathan Jaeger \& Philippe Lambert}

\maketitle


\begin{abstract}
Differential equations (DEs) are commonly used to describe dynamic systems evolving in one (ordinary differential equations or ODEs) or in more than one dimensions (partial differential equations or PDEs). In real data applications the parameters involved in the DE models are usually unknown and need to be estimated from the available measurements together with the state function. In this paper, we present frequentist and Bayesian approaches for the joint estimation of the parameters and of the state functions involved in PDEs. We also propose two strategies to include differential (initial and/or boundary) conditions in the estimation procedure. We evaluate the performances of the proposed strategy on simulated and real data applications.\\

\end{abstract}

\section{Introduction}

Dynamic systems are commonly described by differential equations (DEs). In real data applications the parameters involved in the DE models are usually unknown and need to be estimated from the available measurements together with the state functions. For one dimensional applications this estimation task has been largely discussed in the statistical literature. The most popular procedures rely on nonlinear least squares \citep{biegleretal1986}. These approaches are computationally intensive and often poorly suited for statistical inference. An attractive alternative is represented by the penalized smoothing framework introduced by \citet{ramsayetal2007}. This approach can be viewed as a generalization of the P-spline theory \citep{eilersandmarx1996} where the flexibility acquired by a high dimensional B-spline basis expansion of the state function is counterbalanced by a penalty term expressed using the (set of) ODE(s) that needs to be solved. The fidelity of the final fit to the hypothesized differential model is then tuned using an ODE-adhesion parameter. \citet{jaegerandlambert2013} adapt the latter approach to a full Bayesian framework. This Bayesian alternative offers two major advantages with respect to the frequentist one: the selection of the
ODE-adhesion parameter becomes automatic and uncertainty measures about the parameters can readily be obtained using MCMC.\\
Dealing with dynamics evolving in more than one dimension, partial differential equations (or PDEs) are usually invoked. Inference in models specified by PDEs has not received a large attention in the literature. Here, we discuss statistical approaches to deal with dynamics described by linear PDEs. These kind of equations is used to model large classes of phenomena (e.g. diffusion, heat transfer, price of financial instrument, etc). More precisely, the general form of PDE we consider is given by
\begin{equation}
\mathcal{F}\left(x_{1}, \ldots, x_{p}, u, \frac{\partial u}{\partial x_{1}}, \ldots, \frac{\partial u}{\partial x_{p}}, \frac{\partial ^{2} u}{\partial x_{1} ^ {2}}, \ldots, \frac{\partial ^{2} u}{\partial x_{1} \partial x_{p}}, \dots,\boldsymbol{\theta}\right) = 0,
\label{eq:general_def_pde}
\end{equation}
where $u\left(\boldsymbol{x}\right)$ is the state function evaluated in $\boldsymbol{x} = \left(x_{1},\ldots, x_{p}\right)^{\top}\in \mathbb{R}^{p}$ and $\boldsymbol{\theta}$ is the vector of PDE parameters. In this paper, we consider $\mathcal{F}$ as a linear function in $u$ and its derivatives and with coefficients $\boldsymbol{\theta}$. Examples of this kind of PDE are given in Eqs.~\eqref{eq:pde_simulation} and \eqref{BS_equation} of this paper. In particular, we are interested in estimating the vector of unknown $\boldsymbol{\theta}$ using a set of measurements $\boldsymbol{\zeta} = \zeta\left(\boldsymbol{x}\right) = u\left(\boldsymbol{x}\right) + \boldsymbol{\epsilon}$ where $\boldsymbol{\epsilon}$ can be considered as a vector of independent normally distributed measurement errors.\\
\citet{xunetal2013} propose both frequentist and Bayesian approaches for the estimation of parameters in model driven by PDEs. The frequentist approach should be viewed as a multidimensional generalization of the proposal made by \citet{ramsayetal2007}. The state function solving the differential problem is approximated using tensor product B-spline. The flexibility of the approximation is then counterbalanced by a penalty related to the PDE. In the frequentist proposal, the PDE-penalty is represented by the integral of the PDE operator evaluated at the B-spline approximation of the solution. In their Bayesian approach, they combine a PDE-based penalty with one defined by finite differences (roughness penalty).\\
In this paper we present both frequentist and Bayesian frameworks for the joint estimation of the parameters and the state function involved in PDEs. Our methods generalize those proposed in one-dimensional applications by exploiting a PDE-based penalized tensor product B-spline smoothing approach. Contrary to \citet{xunetal2013}, our Bayesian proposal is directly related to the frequentist one. In Section~\ref{section:without_using_boundaries}, we introduce our estimation procedures. Both frequentist and Bayesian approaches can be adapted in order to include the differential conditions in the estimation process (see Section~\ref{section:including_conditions}). In Section~\ref{section:simulation}, we evaluate the performance of our approach and compare them with the competitors using simulations. In Section~\ref{section:application} we illustrate our proposals dealing with a real data analysis. We conclude the paper with a discussion in Section~\ref{section:discussion}.

\section{Estimation of the unknown differential parameters for unknown differential conditions}
\label{section:without_using_boundaries}

In this section, we introduce the PDE-based penalized tensor product B-spline smoothing approach both in frequentist and Bayesian frameworks without taking the differential conditions (initial and/or boundary) into account.

\subsection{Frequentist approach}

Assume that one observes $\left\{\left(\boldsymbol{\zeta}; \boldsymbol{x}\right) = \left(\zeta_{n}; x_{1, n}, \ldots, x_{p,n}\right); n=1,\ldots,N\right\}$ describing the evolution of the state function $u$ whose dynamics is driven by \eqref{eq:general_def_pde}. Our task is to jointly estimate the state function $u$ and the vector of parameters $\boldsymbol{\theta}$. Following \citet{ramsayetal2007}, we suggest to exploit the DE-based-penalized smoothing approach. More precisely, we denote by $\widetilde{u}\left(\boldsymbol{x}\right)$ the B-spline tensor product approximation of $u\left(\boldsymbol{x}\right)$:
\begin{eqnarray}
\widetilde{u}\left(\boldsymbol{x}\right) & = & \left(\boldsymbol{B_{x_{p}}} \otimes \ldots \otimes \boldsymbol{B_{x_{1}}}\right) \boldsymbol{c} \nonumber \\
                                         & = & \boldsymbol{\mathcal{B}} \boldsymbol{c},
\label{eq:tensor_product}
\end{eqnarray}
where $\boldsymbol{c}$ is the $M^{p}$-vector of spline coefficients and $\boldsymbol{B_{x_{p}}}$ is the $N \times M$ B-spline matrix defined on the $p$-th direction taking a generous numbers of internal knots. The placement of the knots on $x_{p}$-grid can be either regular or not. To simplify the notation, we suppose that the number of B-spline in each direction is the same. The spline coefficients are shrunk by adding a penalty expressed using the PDE model \eqref{eq:general_def_pde}:
\begin{equation}
{PEN}\left(\boldsymbol{c} | \boldsymbol{\theta}\right) = \int{\mathcal{F}\left(\boldsymbol{x}, \widetilde{u}, \boldsymbol{\theta}\right)^{2}d\boldsymbol{x}}.
\label{eq:pde_penalty}
\end{equation}
This penalty term will be close to zero, for given PDE parameters $\boldsymbol{\theta}$, if the approximation to the solution of \eqref{eq:general_def_pde} is consistent with the PDE model. As we are considering linear PDEs, this penalty is just a homogeneous polynomial of degree 2 in the spline coefficients:

\begin{equation*}
{PEN}\left(\boldsymbol{c} | \boldsymbol{\theta}\right) = \boldsymbol{c}^{\top} \boldsymbol{R}\left(\boldsymbol{\theta}\right) \boldsymbol{c} + 2 \boldsymbol{c}^{\top} \boldsymbol{r}\left(\boldsymbol{\theta}\right) + l\left(\boldsymbol{\theta}\right),
\end{equation*}
where $\boldsymbol{R}\left(\boldsymbol{\theta}\right)$ is the penalty matrix, $\boldsymbol{r}\left(\boldsymbol{\theta}\right)$ is the penalty vector and $l\left(\boldsymbol{\theta}\right)$ is a constant not depending on the spline coefficients. A fast computation of these penalty components is given in Appendix~\ref{appendix:elementary_penalty_elements}.\\
The compromise between data fitting and fidelity to the differential model can be assessed using the penalized least square criterion
\begin{equation}
J\left(\boldsymbol{c} | \boldsymbol{\theta}, \gamma, \boldsymbol{\zeta}\right) = \frac{N}{2}\log\left(\tau\right) - \frac{\tau}{2}\left\|\boldsymbol{\zeta} - \widetilde{u}\left(\boldsymbol{x}\right)\right\|^{2} - \frac{\gamma}{2}{PEN}\left(\boldsymbol{c} | \boldsymbol{\theta}\right).
\label{eq:penalized_ls_crit}
\end{equation}
Similarly to \citet{ramsayetal2007}, the estimation process for $\boldsymbol{\theta}$ and $\boldsymbol{c}$ implies on the iteration of two profiling steps. First, for a given value of $\boldsymbol{\theta}$, $\tau$ and $\gamma$, the spline coefficients $\boldsymbol{c}$ are estimated as the maximizer of $J$:
\begin{eqnarray}
\nonumber
\hat{\boldsymbol{c}} & = & \mathop{\mbox{argmax}}_{\boldsymbol{c}} J\left(\boldsymbol{c} | \boldsymbol{\theta}, \tau, \gamma, \boldsymbol{\zeta}\right) \\
                     & = & \left(\tau \boldsymbol{\mathcal{B}}^{\top}\boldsymbol{\mathcal{B}}  + \gamma \boldsymbol{R}\left(\boldsymbol{\theta}\right)\right)^{-1}\left(\tau \boldsymbol{\mathcal{B}}^{\top}\boldsymbol{\zeta} - \gamma \boldsymbol{r}\left(\boldsymbol{\theta}\right)\right).
\end{eqnarray}
Then, given the last available $\hat{\boldsymbol{c}}$, $\boldsymbol{\theta}$ and $\tau$ are estimated by maximizing:
\begin{equation}
\label{eq:precision_theta_est}
H\left(\boldsymbol{\theta}, \tau | \hat{\boldsymbol{c}}, \boldsymbol{\zeta}\right) = \frac{N}{2}\log\left(\tau\right) - \frac{\tau}{2}\left\|\boldsymbol{\zeta} - \boldsymbol{\mathcal{B}} \hat{\boldsymbol{c}}\right\|^{2}.
\end{equation}
The $\gamma$ parameter weights and controls the relative emphasis on goodness-of-fit and solving the partial differential equation. This parameter measures the confidence that one has in the PDE to describe the dynamics in the system. Therefore, we suggest to christen it PDE-adhesion parameter. As usual in penalized smoothing approach, the parameter $\gamma$ has to be selected in a higher optimization level (e.g. using cross-validation, AIC, etc). As suggested by \citet{schall1991}, $\gamma$ can be seen as the ratio of the variance of the residuals and the variance of the penalty. Following this definition, we recommend to use an EM-type procedure to select it. Exact confidence intervals for the PDE parameters cannot easily be obtained. If one is interested in them, we suggest to use bootstrap-type confidence intervals.

\subsection{Bayesian approach}

We propose to adapt the frequentist proposal into a Bayesian framework. Contrary to \citet{xunetal2013}, we will use the same PDE-penalty than in the frequentist approach. The penalty term in Eq.~\eqref{eq:penalized_ls_crit} appears as a term subtracted from the log-likelihood. The log of the joint posterior yields the same fitting criterion using the following model specification:
\begin{equation*}
\left\{\begin{array}{rcl}
   \boldsymbol{\zeta}|\boldsymbol{c},\tau & \sim & \mathcal{N}_{N}\left(\boldsymbol{\mathcal{B}}\boldsymbol{c};\tau^{-1}\boldsymbol{I_{N}}\right),\\
   p(\boldsymbol{c}|\boldsymbol{\theta}, \gamma) & \propto  & \exp\left(- \displaystyle \frac{\gamma}{2}{PEN}\right).
       \end{array} 
\right.
\end{equation*}
The corresponding prior distribution for the spline coefficients is a multivariate normal distribution with mean $\boldsymbol{V_{1}}^{-1}\boldsymbol{v_{1}}$ and variance-covariance $\boldsymbol{V_{1}}^{-1}$ where $\boldsymbol{v_{1}} = - \gamma \boldsymbol{r}\left(\boldsymbol{\theta}\right)$ and $\boldsymbol{V_{1}} = \gamma \boldsymbol{R}\left(\boldsymbol{\theta}\right)$.\\
Further prior distributions have to be given. For the precision of measurement $\tau$, it is convenient to take a gamma distribution $\mathcal{G}\left(a_{\tau}, b_{\tau}\right)$ with mean $a_{\tau}/b_{\tau}$. If no prior information is available on this precision parameter, we recommend either to set $a_{\tau} = b_{\tau}$ equal to a small value (e.g. $10^{-6}$) or to set $a_{\tau}$ equal to one and $b_{\tau}$ to a small value (e.g. $10^{-6}$).\\
The PDE-adhesion parameter appears as a precision parameter in the specification of the prior distribution for the spline coefficients. It is therefore convenient to take a gamma distribution $\mathcal{G}\left(a_{\gamma}, b_{\gamma}\right)$ as prior. In order to put prior confidence on the PDE model, we recommend to set $a_{\gamma}$ to one and $b_{\gamma}$ to a small value (e.g. $10^{-8}$). Indeed, such a prior for $\gamma$ is rather flat although it puts slightly more weight on value of $\log_{10}\left(\gamma\right)$ around $-\log_{10}\left(b_{\gamma}\right)$.\\
For the PDE parameter $\boldsymbol{\theta}$, the prior distribution depends on the problem and is denoted by $p\left(\boldsymbol{\theta}\right)$.\\
The log joint posterior distribution for $\left(\boldsymbol{c}, \boldsymbol{\theta}, \gamma, \tau | \boldsymbol{\zeta}\right)$ can be shown to be:
\begin{equation}
\begin{array}{rcl}
\log\left(p\left(\boldsymbol{c}, \boldsymbol{\theta}, \gamma, \tau | \boldsymbol{\zeta}\right)\right) & = & \displaystyle\frac{N}{2} \log\left(\tau\right) - \displaystyle\frac{\tau}{2} \left\|\boldsymbol{\zeta} - \boldsymbol{\mathcal{B}c} \right\| ^ {2} \\
                                                                                                      & + & \displaystyle\frac{1}{2} \log\left(\mbox{det}\left(\boldsymbol{V}_{1}\right)\right) - \displaystyle\frac{1}{2}\left\{\boldsymbol{c}^{\top}\boldsymbol{V_{1}c} - 2 \boldsymbol{c}^{\top}\boldsymbol{v_{1}} + \boldsymbol{v_{1}}^{\top}\boldsymbol{V_{1}}^{-1}\boldsymbol{v_{1}}\right\} \\
																																																			& + & \left(a_{\tau} - 1\right) \log\left(\tau\right) - b_{\tau}\tau \\
																																																			& + & \left(a_{\gamma} - 1\right) \log\left(\gamma\right) - b_{\gamma} \gamma \\
																																																			& + & \log\left(p\left(\boldsymbol{\theta}\right)\right).
\end{array}
\label{eq:joint_post_dist}
\end{equation}
It can be easily shown that the conditional posterior distribution for the spline coefficients is a multivariate normal distribution with mean $\boldsymbol{V_{2}}^{-1}\boldsymbol{v_{2}}$ and variance-covariance $\boldsymbol{V_{2}}^{-1}$ where $\boldsymbol{V_{2}} = \tau \boldsymbol{\mathcal{B}}^{\top}\boldsymbol{\mathcal{B}} + \boldsymbol{V_{1}}$ and $\boldsymbol{v_{2}} = \tau \boldsymbol{\mathcal{B}}^{\top} \boldsymbol{\zeta} + \boldsymbol{v_{1}}$:
\begin{equation*}
\boldsymbol{c} | \boldsymbol{\theta}, \tau, \gamma, \boldsymbol{\zeta} \sim \mathcal{N}_{M^{p}}\left(\boldsymbol{V_{2}}^{-1}\boldsymbol{v_{2}}; \boldsymbol{V_{2}}^{-1}\right).
\end{equation*}
The conditional posterior distribution for the precision of measurements and for the PDE-adhesion parameter are gamma distributed:
\begin{eqnarray*}
\tau | \boldsymbol{c}, \boldsymbol{\zeta} &\sim& \mathcal{G}\left(\frac{N}{2} + a_{\tau}; \frac{\left\|\boldsymbol{\zeta} - \boldsymbol{\mathcal{B}c} \right\| ^ {2}}{2} + b_{\tau}\right), \\
\gamma | \boldsymbol{\theta}, \boldsymbol{c} &\sim& \mathcal{G}\left(\frac{M^{p}}{2} + a_{\gamma}; \frac{\boldsymbol{c}^{\top}\boldsymbol{R}\left(\boldsymbol{\theta}\right)\boldsymbol{c} + 2 \boldsymbol{c}^{\top}\boldsymbol{r}\left(\boldsymbol{\theta}\right) + \boldsymbol{r}\left(\boldsymbol{\theta}\right)^{\top}\boldsymbol{R}\left(\boldsymbol{\theta}\right)^{-1}\boldsymbol{r}\left(\boldsymbol{\theta}\right)}{2}+ b_{\gamma}\right).
\end{eqnarray*}
The conditional posterior distribution for the PDE parameters $\boldsymbol{\theta}$ is not necessarily of a familiar type. Furthermore, high posterior correlation could occur between the spline coefficients $\boldsymbol{c}$ and the elements of $\boldsymbol{\theta}$. Therefore, we recommend as in \citet{jaegerandlambert2013} to marginalize the joint posterior distribution with respect to the spline coefficients, as given in Appendix~\ref{appendix:log_marg_post_dist}. This marginalization allows us to sample the PDE parameters without the need for updating the spline coefficients during MCMC.

\section{Estimation of the unknown differential parameters including conditions}
\label{section:including_conditions}

In many application, differential conditions arise naturally or are implicitly defined by the observed dynamics. Such information can be included in the statistical framework introduced in Section~\ref{section:without_using_boundaries}.\\
Consider the general partial differential problem given by:
\begin{equation}
\left\{\begin{array}{rcl}
       \mathcal{F}\left(\boldsymbol{x}, u, \boldsymbol{\theta}\right) & = & 0, \\
       \displaystyle\frac{\partial ^ {\left(i\right)}u}{\partial x_{0} ^ {\left(i\right)}} \left(\boldsymbol{x_{0}}\right) & = & v\left(\boldsymbol{x_{0}}\right),
       \end{array}
\right.
\label{eq:general_def_pde_boundaries}
\end{equation}
where $\boldsymbol{x_{0}}$ is a part of the domain where the state function $u$ and/or its $i$th order derivatives are forced to be equal to the function $v$. Using the B-spline approximation, these conditions can be translated into :
\begin{equation}
\boldsymbol{Hc} = v\left(\boldsymbol{x_{0}}\right),
\label{eq:b_spline_representation_boundaries}
\end{equation}
where $\boldsymbol{H}$ is a tensor product B-spline matrix evaluated on the desired support points.
In the frequentist framework, we propose two alternatives in order to introduce the differential conditions in the penalized smoothing approach: least squares and Lagrange multipliers strategies. In the Bayesian settings, we only include the conditions using the least squares strategy.

\subsection{Frequentist approach}

One way to take into account the conditions related to a generic PDE is to consider them as an extra penalty. In the frequentist approach, the fitting criterion $J$ is modified as follows:
\begin{equation}
J\left(\boldsymbol{c} | \boldsymbol{\theta}, \gamma, \kappa, \boldsymbol{\zeta}\right) = \frac{N}{2}\log\left(\tau\right) - \frac{\tau}{2}\left\|\boldsymbol{\zeta} - \widetilde{u}\left(\boldsymbol{x}\right)\right\|^{2} - \frac{\gamma}{2}{PEN}\left(\boldsymbol{\theta} | \boldsymbol{c}\right) - \frac{\kappa}{2}\left(\boldsymbol{Hc} - v\left(\boldsymbol{x_{0}}\right)\right)^{\top}\left(\boldsymbol{Hc} - v\left(\boldsymbol{x_{0}}\right)\right).
\label{eq:penalized_ls_crit_boundaries}
\end{equation}
Note that with such a least square penalty, the approximated state function is not forced to be exactly equal to the conditions. With this fitting criterion, the optimal spline coefficients are obtained:
\begin{eqnarray}
\nonumber
\hat{\boldsymbol{c}} & = & \mathop{\mbox{argmax}}_{\boldsymbol{c}} J\left(\boldsymbol{c} | \boldsymbol{\theta}, \tau, \gamma, \kappa, \boldsymbol{\zeta}\right) \\
                     & = & \left(\tau \boldsymbol{\mathcal{B}}^{\top}\boldsymbol{\mathcal{B}}  + \gamma \boldsymbol{R}\left(\boldsymbol{\theta}\right) + \kappa \boldsymbol{H}^{\top}\boldsymbol{H}\right)^{-1}\left(\tau \boldsymbol{\mathcal{B}}^{\top}\boldsymbol{\zeta} - \gamma \boldsymbol{r}\left(\boldsymbol{\theta}\right) + \kappa \boldsymbol{H}^{\top}v\left(\boldsymbol{x_{0}}\right)\right).
\end{eqnarray}
If $\kappa$ tends to infinity, we simply force the state function to be equal to the differential conditions at the prescribed points. On the other hand, if $\kappa$ is equal to zero, then we just go back to the case considered in Section~\ref{section:without_using_boundaries}. Two alternatives exist to choose the $\kappa$ parameter. The first one consists to fix it to a large value ($10^6$, say). It is a reasonable choice if one is totally confident that the differential conditions in \eqref{eq:general_def_pde_boundaries} should be checked. On the other hand, this parameter can also be selected using standard criteria (such as cross-validation, AIC, etc). The selection of $\lambda$ and $\kappa$ requires to compute these criteria on a grid which can be time consuming. Therefore we advise to fix $\kappa$ to a large value.\\
Using Lagrange multipliers in a frequentist framework, we can force the smoothing function to be exactly consistent with the conditions.\\
The Lagrange function for our constrained maximization problem is:
\begin{equation}
\mathcal{L}\left(\boldsymbol{c}, \boldsymbol{\omega} | \boldsymbol{\theta}, \tau, \gamma, \boldsymbol{\zeta}\right) = \frac{N}{2}\log\left(\tau\right) - \frac{\tau}{2}\left\|\boldsymbol{\zeta} - \widetilde{u}\left(\boldsymbol{x}\right)\right\|^{2} - \frac{\gamma}{2}{PEN}\left(\boldsymbol{\theta} | \boldsymbol{c}\right) - \frac{1}{2}\boldsymbol{\omega}^{\top}\left(\boldsymbol{Hc} - v\left(\boldsymbol{x_{0}}\right)\right),
\label{eq:lagrange_mult_fit_crit}
\end{equation}
where $\boldsymbol{\omega}$ is the vector of Lagrange multipliers. As in \citet{currie2013}, the maximization with respect to $\left(\boldsymbol{c}, \boldsymbol{\omega}\right)$ follows from the solution of:
\begin{equation*}
\left(\begin{array}{c c} \tau \boldsymbol{\mathcal{B}}^{\top}\boldsymbol{\mathcal{B}} + \gamma \boldsymbol{R}\left(\boldsymbol{\theta}\right) & \boldsymbol{H}^{\top} \\
                         \boldsymbol{H} & \boldsymbol{0}
      \end{array}
\right)
\left(\begin{array}{c} \boldsymbol{c} \\
                       \boldsymbol{\omega}
      \end{array}
\right) =
\left(\begin{array}{c} \tau\boldsymbol{\mathcal{B}}^{\top} \boldsymbol{\zeta} - \gamma \boldsymbol{r}\left(\boldsymbol{\theta}\right) \\
                       v\left(\boldsymbol{x_{0}}\right)
      \end{array}
\right).
\end{equation*}
The precision of measurement $\tau$ and the PDE parameters $\boldsymbol{\theta}$ are then estimated in the same way as in Eq.~\eqref{eq:precision_theta_est}. 

\subsection{Bayesian approach}

In a Bayesian setting, the frequentist least squares strategy can be translated into the following prior for the spline coefficients:
\begin{equation*}
p\left(\boldsymbol{c} | \boldsymbol{\theta}, \gamma, \kappa\right) \propto \exp\left(\frac{\gamma}{2}{PEN} - \frac{\kappa}{2}\left(\boldsymbol{Hc} - v\left(\boldsymbol{x_{0}}\right)\right)^{\top}\left(\boldsymbol{Hc} - v\left(\boldsymbol{x_{0}}\right)\right)\right).
\end{equation*}
This prior distribution corresponds to a multivariate normal distribution with mean $\boldsymbol{V_{1}}^{-1}\boldsymbol{v_{1}}$ and variance-covariance $\boldsymbol{V_{1}}^{-1}$ where $\boldsymbol{v_{1}} = - \gamma \boldsymbol{r}\left(\boldsymbol{\theta}\right) + \kappa \boldsymbol{H}^{\top}v\left(\boldsymbol{x_{0}}\right)$ and $\boldsymbol{V_{1}} = \gamma \boldsymbol{R}\left(\boldsymbol{\theta}\right) + \kappa \boldsymbol{H}^{\top}\boldsymbol{H}$.\\
As in the frequentist approach, two alternatives are available for $\kappa$. One can either fix it to a large value or consider it as random with a gamma prior distribution $\mathcal{G}\left(a_{\kappa}; b_{\kappa}\right)$ where $a_{\kappa}$ is equal to one and $b_{\kappa}$ is equal to a small value (e.g. $10^{-6}$). This choice for the prior distribution for $\kappa$ translates prior confidence in the differential conditions. The log joint posterior distribution, the log marginalized posterior distribution and all the conditional posterior distributions can be found in Appendix~\ref{appendix:post_dist_least_squares}.
For the Lagrange multipliers strategy, the possible Bayesian translation would require to project the spline coefficients on a parameter sub-space where the constraints are met: we have no practical implementation of such a strategy for the moment.

\section{Simulation}
\label{section:simulation}

An intensive simulation study was set up to study the properties of the proposed estimation strategies. We consider the following diffusion partial differential equation:
\begin{equation}
\label{eq:pde_simulation}
\mathcal{F}\left(x_{1}, x_{2}, u, \frac{\partial u}{\partial x_{1}}, \frac{\partial u}{\partial x_{2}}, \theta_{1}, \theta_{2}\right) = u_{x_1} + \theta_{1} u_{x_2} + \theta_{2} u = 0,\ \mbox{with}\ x_{1} \in \mathbb{R},\ x_{2} \in \mathbb{R}^{+},
\end{equation}
where $u_{x_{i}}$ denotes the first derivative of the state function with respect to $x_{i}$. For simulation purposes, we take $\theta_{1} = 0.5$ and $\theta_{2} = 1.5$ and consider the condition $ u(x_{1}, 0) = \displaystyle \frac{1}{1 + x_{1}^{2}}$. With these settings the considered differential problem has closed form solution:
\[ 
u(x_{1}, x_{2}) = \frac{\exp(- 3 x_{2})}{1 + 4 x_{2}^{2} - 4 x_{1} x_{2} + x_{1}^{2}}.
\]
This expression contaminated with gaussian noise will be used to generate the data. The estimation abilities of our proposals will be compared to the methods presented by \citet{xunetal2013} (see Section~\ref{subsection:comparison}).

\subsection{Simulation using the PDE-P-splines approaches}
\label{subsection:simulation_using_pde_p_splines_approach}

The data used for the simulations of this section have been obtained by adding a Gaussian noise component to the analytic solution computed on the grid of equidistant points $x_{1, i} \in [-3, 3],\ x_{2, i} \in [0, 1]$ with $i = 1,...,2500$.
Different levels for the precision of measurements $\tau = 10000, 400, 100$ have been used to simulate the data in such a way to obtain a low, medium and large level of noise in the measurements. Figure~\ref{fig_simulation_settings} shows three possible simulated data clouds.
\begin{figure}
\animategraphics[width = 1\linewidth, loop, autoplay]{1.5}{movie_simulation_all}{3}{3}
\caption{Three simulated data sets (dots) obtained considering a diffusion dynamics and different levels for the precision of measurements. The surfaces represent the signal used in the simulation.}
\label{fig_simulation_settings}
\end{figure}
We are interested in evaluating the efficiency of the proposed estimation procedures. For this reason we consider 500 simulated datasets per noise level and compute, for the frequentist approaches, the relative bias (in percent), the relative root mean squared error and the standard deviation of the estimated differential equation parameters together with the precision of measurements and the optimal adhesion parameter. On the other hand, with the Bayesian approaches, we compute the relative bias (based on the posterior mean), the relative root mean squared error (based on the posterior mean), the mean posterior standard deviation and the $80\%$ and $95\%$ coverage probabilities (based on the HPD intervals). \\
In this simulation study we stress our proposal with and without taking into account the differential conditions. For the frequentist procedures we include the differential conditions using either the least squares or the Lagrange multiplier strategies. For the Bayesian approach, we only investigate the least squares strategy as it enables to marginalize the joint posterior distribution with respect to the spline coefficients.\\
In order to decrease the computation burden, we consider in this simulation study unequidistant knots. We use a set of 28 spline basis in the $x_{1}$ direction and 13 spline basis in the $x_{2}$ direction. In both direction, more knots have been located in the part of the domain where variation of the signal appears more evident. The selection of knots location is not mandatory and fine grid of equidistant knots produces equivalent estimate but requires higher computational efforts. For the degree of the B-splines we found that a useful rule of thumb is to set it equal to the degree of the differential equation plus two.

\subsubsection{Simulation results without considering differential conditions}
\label{subsubsection:simulation_noboundary}

In Table~\ref{table_simulation_noboundary} the results of the simulation study described above are shown. In this case the differential conditions have not been taken into account in the estimation process. The cited table shows some interesting features. The relative bias (in percent) for the estimated parameters is small both for the frequentist and Bayesian approaches. On the other hand, looking at the estimated PDE parameters, the relative bias obtained within the Bayesian framework is positive and larger (but still very small) than for the frequentist estimates. This could be explained by the choice of the posterior mean for the point estimate while the estimates for the precision parameter look similar for the two approaches. The RMSEs seem to be larger for the PDE parameters in the Bayesian approach and larger for the precision of measurement in the frequentist approach. This can be explained looking at the bias. On the other hand, based on the STDs, the estimation procedure seems quite robust against the noise level. For the Bayesian estimates, the mean posterior standard deviations tends to increase with the level of the noise for the PDE parameters and to decrease when the noise level increases for the $\tau$  parameter. Only some estimated coverage probabilities (at 80\% and 95\% level) are in agreement with the nominal values. On the other hand, for a high level of noise, there is evidence of an under coverage effect that could be explained by the fact that we are estimating the differential conditions without considering them. Finally, the (average) optimal adhesion parameters $\gamma$ shown at the bottom of each sub-table are fairly large indicating a strong (posterior) confidence in the appropriateness of the differential model used in the penalty to describe the observed dynamics. Figure~\ref{fig_average_simulation_profiles} shows three possible smoothing surfaces coming from this simulation study. As expected, if we don't include conditions in the estimation procedure, the fitted surface tends to be sensitive to noise variability showing wiggly part for large values of $x_{1}$ and moderate value of $x_{2}$ (see first column of Figure~\ref{fig_average_simulation_profiles}).
\begin{table}[htbp]
  \centering
  \caption{Frequentist and Bayesian simulation for diffusion PDE smoothing example without considering the differential conditions. At the end of each section in the table the average optimal adhesion parameters are listed.}
    \begin{tabular}{rrl|ccc}
		\cline{4-6}
          & \multicolumn{2}{r}{}  & \multicolumn{3}{c}{\textbf{True parameter values}} \\
          & \multicolumn{2}{r}{}  & $\theta_1 = 0.5$ & $\theta_2 = 1.5$ & $\tau = 1/\sigma^{2}$ \\
\cline{2-6}  \\   \multicolumn{1}{c}{\multirow{12}[2]{*}{\begin{sideways}\textbf{Frequentist}\end{sideways}}} &
                                    &$\sigma^2 = 0.01^2 $& 0.08\% & 0.09\% & 2.55\% \\
    \multicolumn{1}{c}{} & R-BIAS   &$\sigma^2 = 0.05^2 $& 0.45\% & 0.47\% & 2.22\% \\
    \multicolumn{1}{c}{} &          &$\sigma^2 = 0.1^2  $& 1.16\% & 1.12\% & 6.48\% \\
    \multicolumn{1}{c}{} &          &       &       &  \\
    \multicolumn{1}{c}{} &          &$\sigma^2 = 0.01^2 $& 4.49E-03 & 5.00E-03 & 3.91E-02 \\
    \multicolumn{1}{c}{} & R-RMSE   &$\sigma^2 = 0.05^2 $& 2.27E-02 & 2.52E-02 & 3.85E-02 \\
    \multicolumn{1}{c}{} &          &$\sigma^2 = 0.1^2  $& 4.67E-02 & 6.81E-01 & 2.99E-02 \\
    \multicolumn{1}{c}{} &          &       &       &  \\
    \multicolumn{1}{c}{} &          &$\sigma^2 = 0.01^2 $& 4.42E-03 & 4.93E-03 & 2.98E-02 \\
    \multicolumn{1}{c}{} & R-STD    &$\sigma^2 = 0.05^2 $& 2.23E-02 & 2.48E-02 & 3.15E-02 \\
    \multicolumn{1}{c}{} &          &$\sigma^2 = 0.1^2  $& 4.53E-02 & 5.04E-02 & 2.92E-02 \\
		&&\multicolumn{3}{c}{ $E(\gamma)= \{\mbox{\textit{2.36E+08, 5.20E+07,	4.30E+07}} \}$ } &\\
\cline{2-6} \\  \multicolumn{1}{c}{\multirow{16}[2]{*}{\begin{sideways}\textbf{Bayesian}\end{sideways}}} &
                                    &$\sigma^2 = 0.01^2 $& 0.06\% & 0.07\% & 0.33\% \\
    \multicolumn{1}{c}{} & R-BIAS   &$\sigma^2 = 0.05^2 $& 1.14\% & 1.35\% & 0.15\% \\
    \multicolumn{1}{c}{} &          &$\sigma^2 = 0.1^2  $& 4.02\% & 4.79\% & 0.09\% \\
    \multicolumn{1}{c}{} &          &       &       &  \\
    \multicolumn{1}{c}{} &          &$\sigma^2 = 0.01^2 $& 4.46E-03 & 4.96E-03 & 2.93E-02 \\
    \multicolumn{1}{c}{} & R-RMSE   &$\sigma^2 = 0.05^2 $& 2.54E-02 & 2.86E-02 & 2.90E-02 \\
    \multicolumn{1}{c}{} &          &$\sigma^2 = 0.1^2  $& 6.27E-02 & 7.20E-02 & 2.90E-02 \\
    \multicolumn{1}{c}{} &          &       &       &  \\
    \multicolumn{1}{c}{} &          &$\sigma^2 = 0.01^2 $& 2.19E-03 & 7.31E-03 & 2.85E+02 \\
    \multicolumn{1}{c}{} & MPSD     &$\sigma^2 = 0.05^2 $& 1.10E-02 & 3.66E-02 & 1.14E+01 \\
    \multicolumn{1}{c}{} &          &$\sigma^2 = 0.1^2  $& 2.35E-02 & 7.84E-02 & 2.84E+00 \\
    \multicolumn{1}{c}{} &          &       &       &  \\
    \multicolumn{1}{c}{} &          &$\sigma^2 = 0.01^2 $& \textbf{80.2}  & \textbf{80.6}  & \textbf{80.6} \\
    \multicolumn{1}{c}{} & CP80     &$\sigma^2 = 0.05^2 $& \textbf{76.6}  & 73.4  & \textbf{80.0} \\
    \multicolumn{1}{c}{} &          &$\sigma^2 = 0.1^2  $& 69.4  & 68.6  & \textbf{79.6} \\
    \multicolumn{1}{c}{} &          &       &       &  \\
    \multicolumn{1}{c}{} &          &$\sigma^2 = 0.01^2 $& \textbf{94.4} & \textbf{93.4} & 92.6 \\
    \multicolumn{1}{c}{} & CP95     &$\sigma^2 = 0.05^2 $& 91.0  & 93.0  & 92.2 \\
    \multicolumn{1}{c}{} &          &$\sigma^2 = 0.1^2  $& 89.2 & 87.0  & \textbf{93.2} \\
		&&\multicolumn{3}{c}{$E(\gamma)= \{\mbox{\textit{1.48E+07,	2.02E+07,	3.01E+07}} \} $ }&\\
		\cline{2-6}    \end{tabular}%
  \label{table_simulation_noboundary}%
\end{table}%

\subsubsection{Simulation results considering the differential conditions}
\label{subsubsection:simulation_boundary}
\begin{figure}
\includegraphics[width = 1\linewidth]{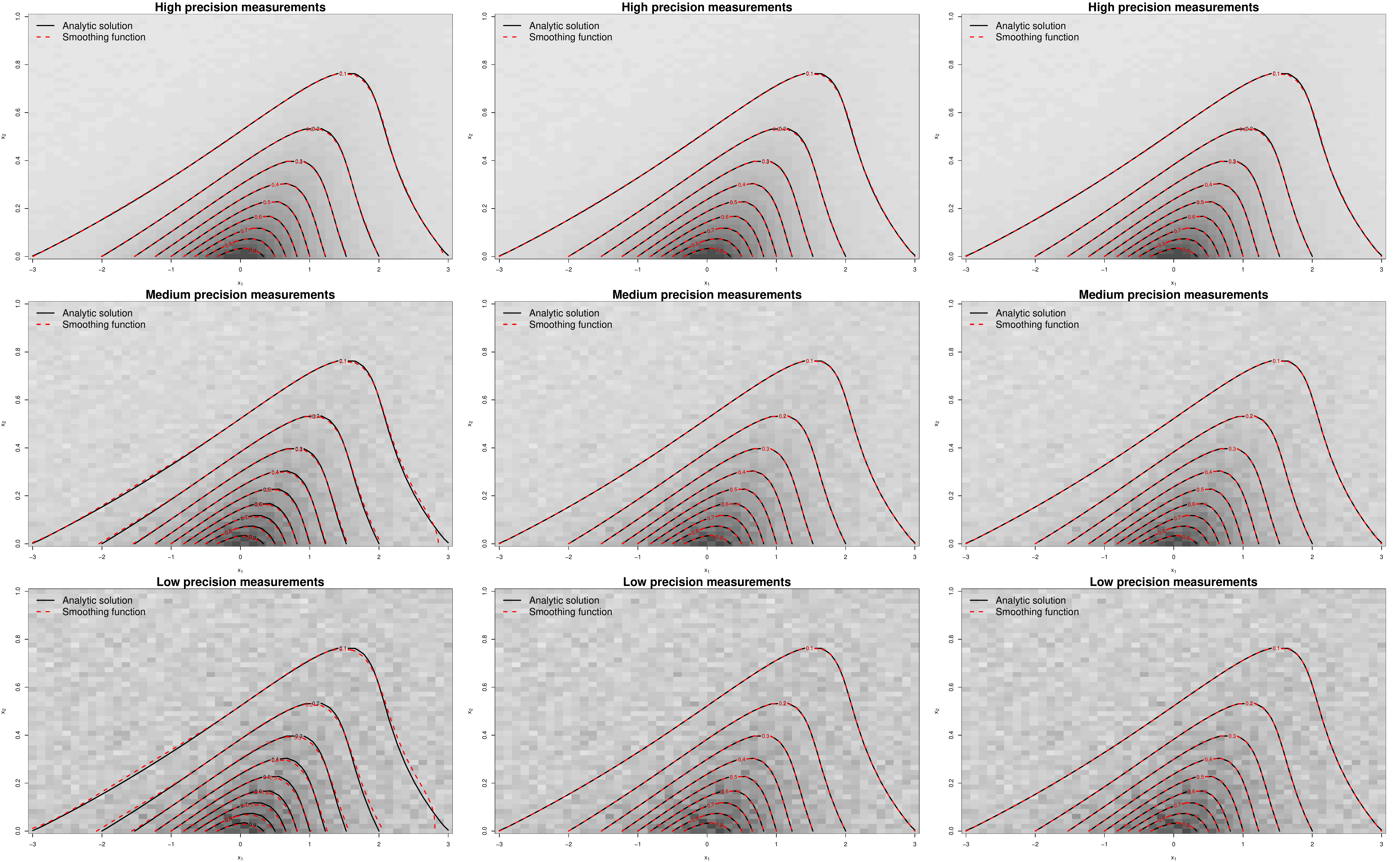}
\caption{Typical profiles for the frequentist approaches. The first column shows the smoothing surfaces estimated for unknown conditions. The middle column shows the smoothing surfaces estimated using the differential conditions imposed by least squares, while the third column shows the smoothing surface using the Lagrange multipliers approach. The gray background indicates the simulated observations. Darker is the background, larger is the corresponding observed value on the vertical axis (the value of $\boldsymbol{\zeta}$).}
\label{fig_average_simulation_profiles}
\end{figure}

The results of the simulation study taking into account the conditions using least squares (both frequentist and Bayesian approaches with fixed $\kappa = 10^{6}$) or Lagrange multipliers (only frequentist approach) are shown in Table~\ref{table_simulation_boundary}. It appears that the relative bias for the estimated parameters is small both for the frequentist and Bayesian approaches. One can notice that the bias decreases in absolute value when the differential conditions are introduced in the estimation procedure in the case of moderate and high noise variability. As before, for the frequentist estimates, the RMSEs seem to increase together with the level of measurement noise. On the other hand, based on the STDs, the estimation procedure seems quite robust against the noise level. The estimated standard deviations for the PDE parameters seem to be lower than what we observed without considering the conditions. Looking at the frequentist estimates the results obtained imposing conditions by least squares and using Lagrange multipliers look really similar. For the Bayesian estimates, the mean posterior standard deviations tends to increase with the level of the noise for the PDE parameters and to decrease when the noise level increases for the $\tau$  parameter. As expected, including the differential condition, the mean standard deviation for the PDE parameters tends to decrease. The estimate coverage probabilities are in agreement with the nominal values in almost all the simulation settings. This is probably due to the fact that the conditions are explicitly specified and not estimated. The average optimal adhesion parameters $\gamma$ given in each sub-tables, indicate a stronger (posterior) confidence in the PDE model as descriptor of the dynamics of the observed state.
The middle and the third column of Figure~\ref{fig_average_simulation_profiles} show three smoothing surfaces extracted from the simulation runs. Compared to the first column, the inclusion of the differential conditions forces the smoothing surface to be adherent to the closed form solution over the entire domain even for lower precision of the measurements.
\begin{table}[htbp]
  \centering
  \caption{Frequentist and Bayesian simulation for diffusion PDE smoothing example considering the differential conditions. At the end of each section in the table the average optimal adhesion parameters are listed.}
    \begin{tabular}{rrl|ccc}
		\cline{4-6}
          & \multicolumn{2}{r}{}  & \multicolumn{3}{c}{\textbf{True parameter values}} \\
          & \multicolumn{2}{r}{} & $\theta_1 = 0.5$ & $\theta_2 = 1.5$ & $\tau = 1/\sigma^{2}$ \\
\cline{2-6}  \\   \multicolumn{1}{c}{\multirow{12}[2]{*}{\begin{sideways}\textbf{Frequentist (least squares)}\end{sideways}}} &
                                    &$\sigma^2 = 0.01^2 $& 0.11\% & 0.07\% & 1.51\% \\
    \multicolumn{1}{c}{} & R-BIAS   &$\sigma^2 = 0.05^2 $& 0.15\% & 0.05\% & 2.12\% \\
    \multicolumn{1}{c}{} &          &$\sigma^2 = 0.1^2  $& 0.16\% & 0.02\% & 0.58\% \\
    \multicolumn{1}{c}{} &          &       &       &  \\
    \multicolumn{1}{c}{} &          &$\sigma^2 = 0.01^2 $& 3.56E-03 & 3.83E-03 & 6.20E-02 \\
    \multicolumn{1}{c}{} & R-RMSE   &$\sigma^2 = 0.05^2 $& 1.62E-02 & 1.77E-02 & 3.69E-02 \\
    \multicolumn{1}{c}{} &          &$\sigma^2 = 0.1^2  $& 3.25E-02 & 6.68E-01 & 2.98E-02 \\
    \multicolumn{1}{c}{} &          &       &       &  \\
    \multicolumn{1}{c}{} &          &$\sigma^2 = 0.01^2 $& 3.40E-03 & 3.77E-03 & 6.02E-02 \\
    \multicolumn{1}{c}{} & R-STD    &$\sigma^2 = 0.05^2 $& 1.61E-02 & 1.77E-02 & 3.03E-02 \\
    \multicolumn{1}{c}{} &         &$\sigma^2 = 0.1^2  $& 3.25E-02 & 3.57E-02 & 2.92E-02 \\
		&&\multicolumn{3}{c}{ $E(\gamma)= \{\mbox{\textit{2.10E+08, 2.94E+07,	1.21E+07}} \}  $} &\\
		\cline{2-6} \\ \multirow{12}[2]{*}{\begin{sideways}\textbf{Frequentist (Lagrange)}\end{sideways}} &
					                          &$\sigma^2 = 0.01^2 $& 0.11\% & 0.06\% & 2.24\% \\
    \multicolumn{1}{c}{} & R-BIAS   &$\sigma^2 = 0.05^2 $& 0.14\% & 0.05\% & 2.13\% \\
    \multicolumn{1}{c}{} &          &$\sigma^2 = 0.1^2  $& 0.16\% & 0.01\% & 0.58\% \\
    \multicolumn{1}{c}{} &          &       &       &  \\
    \multicolumn{1}{c}{} &          &$\sigma^2 = 0.01^2 $& 3.40E-03 & 3.58E-03 & 3.72E-02 \\
    \multicolumn{1}{c}{} & R-RMSE   &$\sigma^2 = 0.05^2 $& 1.62E-02 & 1.77E-02 & 3.70E-02 \\
    \multicolumn{1}{c}{} &          &$\sigma^2 = 0.1^2  $& 3.25E-02 & 6.68E-01 & 2.98E-02 \\
    \multicolumn{1}{c}{} &          &       &       &  \\
    \multicolumn{1}{c}{} &          &$\sigma^2 = 0.01^2 $& 3.22E-03 & 3.53E-03 & 2.97E-02 \\
    \multicolumn{1}{c}{} & R-STD    &$\sigma^2 = 0.05^2 $& 1.62E-02 & 1.77E-02 & 3.03E-02 \\
    \multicolumn{1}{c}{} &          &$\sigma^2 = 0.1^2  $& 3.25E-02 & 3.57E-02 & 2.93E-02 \\
		&&\multicolumn{3}{c}{ $E(\gamma)= \{\mbox{\textit{2.11E+08, 3.62E+07,	2.08E+07}} \}  $} &\\
\cline{2-6} \\  \multicolumn{1}{c}{\multirow{16}[2]{*}{\begin{sideways}\textbf{Bayesian}\end{sideways}}} &
                                    &$\sigma^2 = 0.01^2 $& 0.02\% & 0.01\% & 0.28\% \\
    \multicolumn{1}{c}{} & R-BIAS   &$\sigma^2 = 0.05^2 $& 0.24\% & 0.22\% & 0.23\% \\
    \multicolumn{1}{c}{} &          &$\sigma^2 = 0.1^2  $& 0.81\% & 0.82\% & 0.23\% \\
    \multicolumn{1}{c}{} &          &       &       &  \\
    \multicolumn{1}{c}{} &          &$\sigma^2 = 0.01^2 $& 3.22E-03 & 3.54E-03 & 2.91E-02 \\
    \multicolumn{1}{c}{} & R-RMSE   &$\sigma^2 = 0.05^2 $& 1.64E-02 & 1.79E-02 & 2.90E-02 \\
    \multicolumn{1}{c}{} &          &$\sigma^2 = 0.1^2  $& 3.39E-02 & 3.71E-02 & 2.91E-02 \\
    \multicolumn{1}{c}{} &          &       &       &  \\
    \multicolumn{1}{c}{} &          &$\sigma^2 = 0.01^2 $& 1.61E-03 & 5.38E-03 & 2.84E+02 \\
    \multicolumn{1}{c}{} & MPSD     &$\sigma^2 = 0.05^2 $& 7.70E-03 & 2.57E-02 & 1.14E+01 \\
    \multicolumn{1}{c}{} &          &$\sigma^2 = 0.1^2  $& 1.56E-02 & 5.21E-02 & 2.84E+00 \\
    \multicolumn{1}{c}{} &          &       &       &  \\
    \multicolumn{1}{c}{} &          &$\sigma^2 = 0.01^2 $& \textbf{80.2}  & \textbf{80.2}  & \textbf{80.6} \\
    \multicolumn{1}{c}{} & CP80     &$\sigma^2 = 0.05^2 $& \textbf{78.4}  & \textbf{78.2}  & \textbf{80.6} \\
    \multicolumn{1}{c}{} &          &$\sigma^2 = 0.1^2  $& \textbf{78.0}  & \textbf{79.0}  & \textbf{79.8} \\
    \multicolumn{1}{c}{} &          &       &       &  \\
    \multicolumn{1}{c}{} &          &$\sigma^2 = 0.01^2 $& \textbf{95.0} & \textbf{94.2} & \textbf{93.2} \\
    \multicolumn{1}{c}{} & CP95     &$\sigma^2 = 0.05^2 $& 92.4 & \textbf{93.4} & 93.0 \\
    \multicolumn{1}{c}{} &          &$\sigma^2 = 0.1^2  $& \textbf{93.2} & \textbf{93.2} & 93.0 \\
		&&\multicolumn{3}{c}{ $E(\gamma)= \{\mbox{\textit{1.05E+08,	1.48E+08,	1.37E+08}} \} $}&\\
		\cline{2-6}    \end{tabular}%
  \label{table_simulation_boundary}%
\end{table}%

\subsection{Comparison of the PDE-P-splines approaches with the one by \citet{xunetal2013}}
\label{subsection:comparison}

\begin{figure}
\centering
\includegraphics[width = 1.15\linewidth, angle = 90]{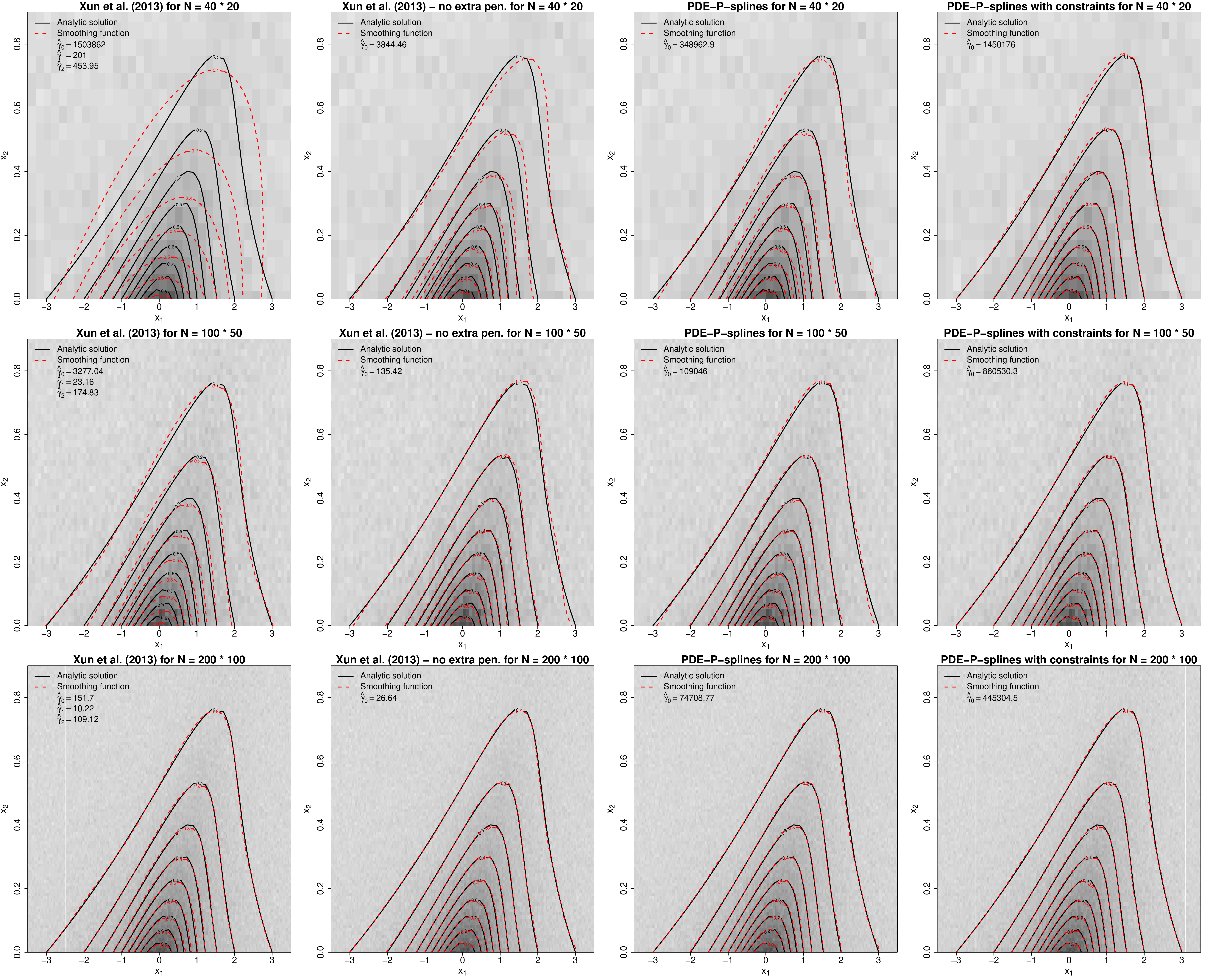}
\caption{Posterior pointwise smoothing surfaces estimated with \citet{xunetal2013} approach with and without the extra finite difference penalty (first and second rows) and smoothing surfaces estimated using the PDE-P-splines approaches without and with the boundary conditions imposed by least squares (third and fourth rows). In each column, the surfaces estimated for different sample sizes are depicted ($n = 40 \times 20$ in the left column, $n = 100 \times 50$ in the middle column and $n = 200 \times 100$ in the right column). The gray background indicates the simulated observations. Darker is the background, larger is the corresponding observed value on the vertical axis (the value of $\boldsymbol{\zeta}$).}
\label{fig_comparison_surface}
\end{figure}

\begin{figure}
\includegraphics[width = 1\linewidth]{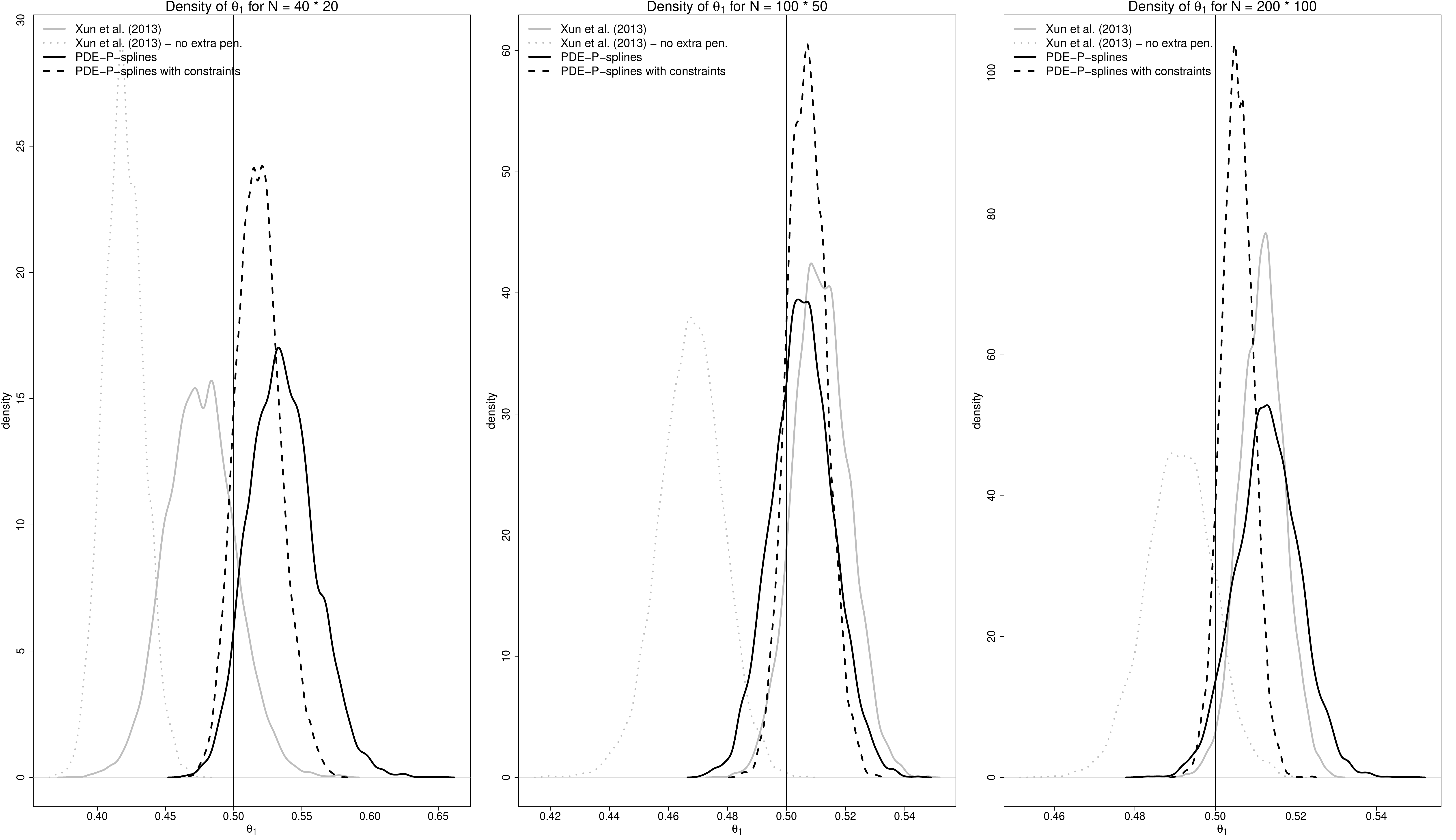}
\includegraphics[width = 1\linewidth]{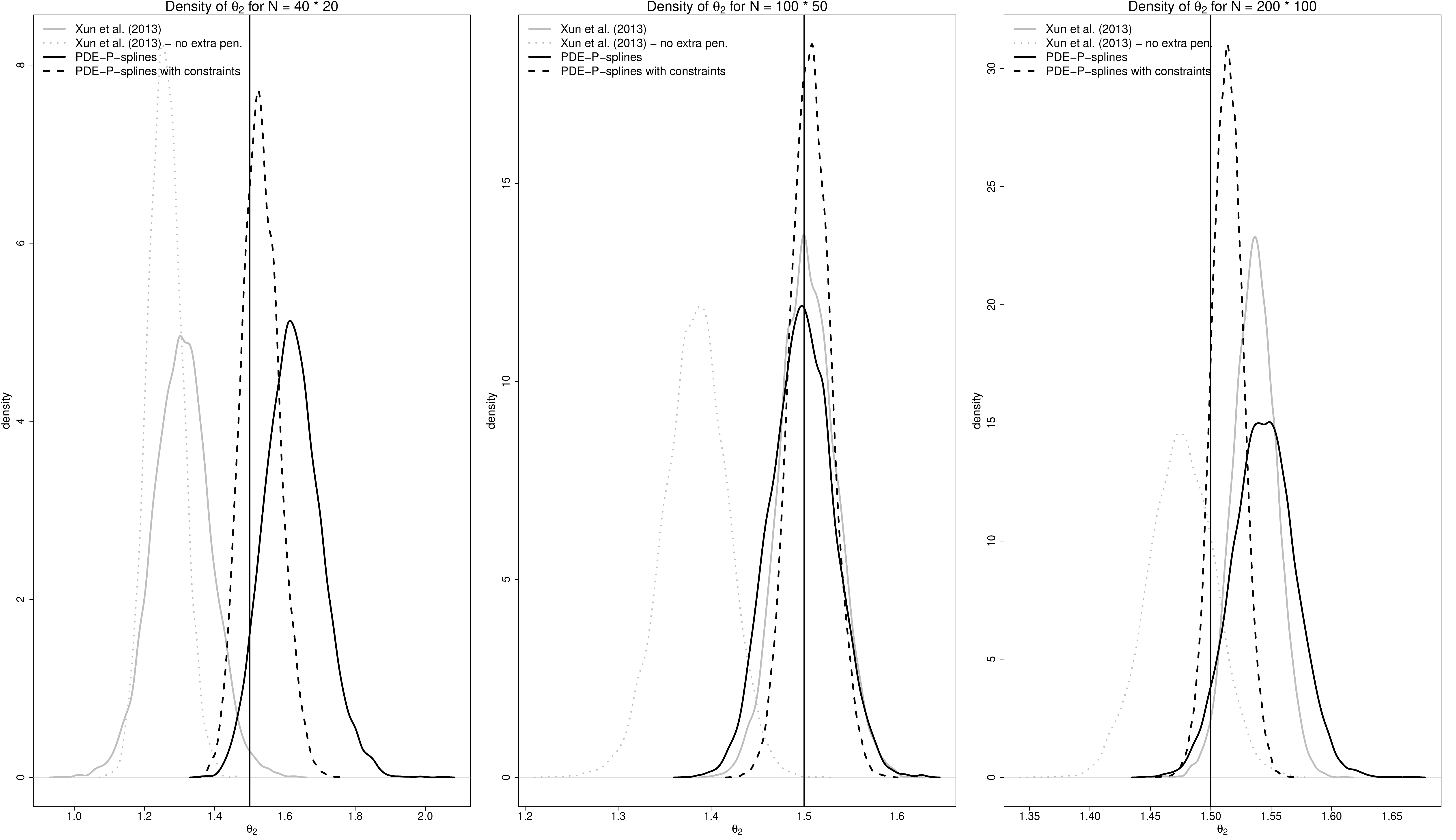}
\caption{Posterior densities for $\theta_{1} = 0.5$ and $\theta_{2} = 1.5$, respectively. Densities in gray are those obtained using the approaches proposed by \citet{xunetal2013} (solid curve with extra finite difference penalty and dotted curve without it). Densities in black are those obtained using the PDE-P-splines approaches (solid curve without the differential condition and dashed one with it). In each column, the posterior densities for different sample sizes are depicted ($n = 40 \times 20$ in the left column, $n = 100 \times 50$ in the middle column and $n = 200 \times 100$ in the right column).}
\label{fig_post_density_thetas}
\end{figure}

In the Bayesian framework, \citet{xunetal2013} proposes to use a penalty that combines the requirement for PDE adhesion (in a $L^{2}$ sense) and the one for the smoothness of the estimated surface (using extra finite difference penalties). Our interest is to compare this proposal with our Bayesian approaches. In order to underline the weaknesses of the PDE-based penalty used in \citet{xunetal2013}, we also consider the case where the extra finite difference penalties are not included in the model specification. The two main differences between our proposals and the one by \citet{xunetal2013} lay in the construction of the penalty term and in computation of the appropriate constant of normalization for the prior distribution for the splines coefficients. For the construction of the penalty term, we used an integrated penalty that ensure the fidelity of the B-spline approximation to the proposed PDE on the entire domain whereas \citet{xunetal2013} build the penalty only on the observed points. Note that our Bayesian approach is a direct ``Bayesianization'' of the frequentist ones. As appears in Eq.~\eqref{eq:joint_post_dist}, we compute the constant of normalization of the prior distribution for the spline coefficients as the determinant of the PDE-penalty matrix whereas \citet{xunetal2013} just consider a power of the smoothing parameters as constant of normalization for that prior.\\
The data used for the comparison have been obtained by adding a Gaussian noise component ($\tau = 400$) to the analytic solution of the PDE given in Eq.~\eqref{eq:pde_simulation} computed on the grid of equidistant points $x_{1, i} \in [-4, 4],\ x_{2, i} \in [0, 1]$ with $i = 1,...,n$, where $n = 40 \times 20, 100 \times 50, 200 \times 100$ (low, medium and large sample sizes).\\
Figure~\ref{fig_comparison_surface} shows the estimated surfaces obtained using the four approaches. For the large sample sizes, all the estimated smoothing surfaces appear adherent to the state function, except for the approach of \citet{xunetal2013} that seems less efficient in reproducing the peak of the signal. When the sample size decreases, our approaches seem to perform better that the competitors in recovering the underlying signal. In the case where the sample is limited, the approach of \citet{xunetal2013} tends to oversmooth the data and seems to put not enough weight on the PDE-model penalty. On the other hand, for the same simulation settings, the modified version of the model by \citet{xunetal2013} without the extra finite difference penalty seems to better catch the signal. The posterior densities for the PDE parameters in the modified \citeauthor{xunetal2013} approach appears clearly left shifted (see Figure~\ref{fig_post_density_thetas}). The extra finite difference penalty seems to mitigate the introduced bias. The complete constant of normalization in the prior distribution for the spline coefficients centers the posterior distributions around the true PDE parameters. Finally, whatever the sample size, the variability of the posteriors decreases when the differential conditions are introduced in the model.

\section{Application}
\label{section:application}

\begin{figure}
\centering
\includegraphics[width = 1\linewidth]{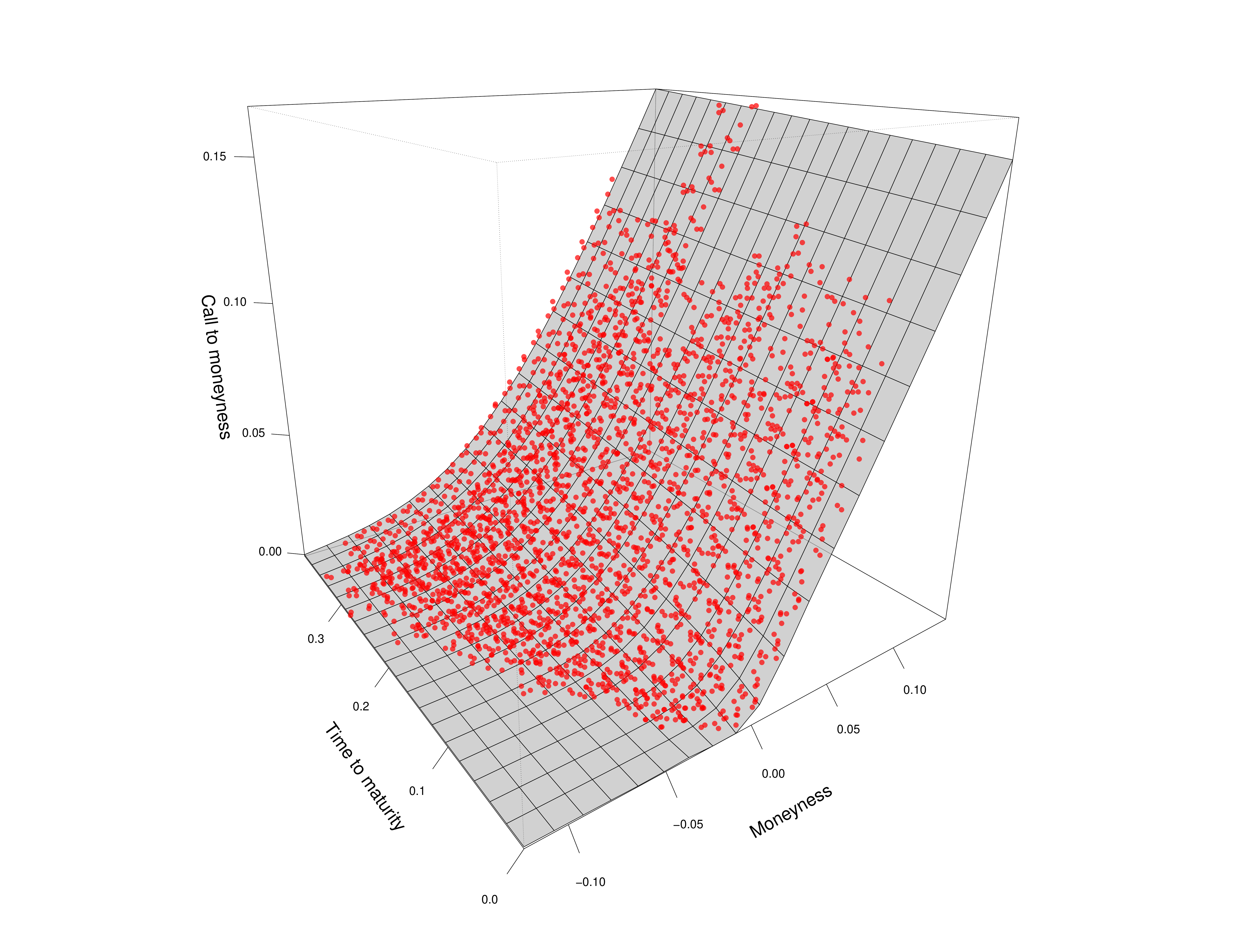}
\caption{Observations (red dots) and smoothing surface obtained applying a constrained PDE-P-spline estimation procedure to the SP-500 European call option prices analyzed by \citet*{carmona2004}. The prices in the plot are scaled by the observed strikes and the moneyness is defined as the log of the ratio between the observed spot and strike prices. The estimated implied volatility was $\hat{\sigma} = 1.003E-01$ with an optimal adhesion parameter equal to $\hat{\gamma} = 47385275$.}
\label{fig_BS_estimates}
\end{figure}

In this section we evaluate the performance of the proposed approaches dealing with a real data application. We consider as example the analysis of a series of option prices traded for different strike prices and maturities. Quoting \citet*{cox1979}, an option is a contract that gives the right, but not the obligation, to buy or sell a risky asset at a predetermined fixed price within (American style) or at (European style) a given date (maturity of the contract). This kind of financial instruments allow to bet about the future evolution of the price of the underlying asset that can be another contract (a stock for example), a market index or a commodity. Options and other related instruments are often defined ``derivatives'' given that their price is derived from the market value ($S$) of the underlying risky asset. \\
Two different kinds of option contracts are distinguished according to the right that they give to the holder. ``Call'' options give the right to buy the underlying asset while ``put'' give the right to sell it. These characteristics determine the payoff functions at maturity ($T$):
\begin{equation*}
C(S, T) = \left\{
   \begin{array}{ll}
      0\  & \mbox{if}\ S_{T} \leq E \\
      S_{T} - E\ & \mbox{if}\ S_{T} > E,
   \end{array}
 \right.
\end{equation*}
\begin{equation*}
P(S, T) = \left\{
   \begin{array}{ll}
      0\  &\mbox{if}\ S_{T} \geq E \\
      E-S_{T}\ & \mbox{if}\ S_{T} < E,
   \end{array}
 \right.
\end{equation*}
where $C$ and $P$ indicate call and put prices respectively, $E$ is the strike price and $S_{T}$ represents the market price of the underlying asset at maturity. 
The current price of the contract is then computed as the present value of the payoff at maturity. For this reason it is necessary to describe the dynamics of the underlying asset in order to price the derivative contract ``written'' on it. \\
The \citet{black1973} framework is the most famous pricing model for financial derivatives. The B\&S model comes from the assumption that the price of the underlying risky asset follows a geometric Brownian motion in the instantaneous time with a constant standard deviation (volatility). This assumption leads to the well known second order partial differential equation:
\begin{equation}\label{BS_equation}
\mathcal{F}\left(S, t, C, \frac{\partial C}{\partial S}, \frac{\partial C}{\partial t}, \frac{\partial^{2} C}{\partial S^{2}}, \sigma\right) = \frac{\partial C}{\partial t} + r S \frac{\partial C}{\partial S} + \frac{1}{2} \sigma^{2} S^{2} \frac{\partial^{2} C}{\partial S^{2}} - r C = 0,
\end{equation}
where the parameter $\sigma$ identifies the ``implied volatility'' and $r$ is the observed risk free rate. For European options (exercise allowed only at a given date) a close form solution can be found introducing the following no-arbitrage terminal and boundary conditions:
\begin{eqnarray}\label{BS_conditions}
\nonumber
&& C(0, t) = 0,\\
&& C(S, t) \approx S \ \mbox{for}\ S \rightarrow \infty,\\
\nonumber
&& C(S, T) = \max(S-E, 0) = (S-E)^{+}.
\end{eqnarray}
This framework provides a concise and easy to interpret description of the price dynamics. Furthermore, this model allows to compute the theoretical option prices for unobserved strikes and maturities and, inverting the solution of the PDE above, to recover the volatility ($\sigma$) implied in the observed prices (the volatility parameter that solves the equation for the observed prices). Moving from these considerations, we suggest to achieve these goals including the PDE in~\eqref{BS_equation} and the no-arbitrage differential conditions in~\eqref{BS_conditions} in a PDE-P-spline approach. Note that in this application we share with the theoretical model the unrealistic assumption of constant volatility over strikes and maturities. Besides the limitations implicit in the B\&S model, we believe that the proposed estimation procedure is flexible enough to ensure appropriate estimates even if the hypotheses of the differential model do not completely match with the data evidence.  \\
We conduct our analysis using the data proposed by \citet*{carmona2004}. The dataset counts 2800 prices of European call options written on the SP-500 index and traded in 1993 with expiration between 1993 and 1994. The six variables included in the dataset indicate the value of the index, the strike prices of the option, the time to expiration (in fraction of year), the spot interest rate, the observed implied volatilities and the prices of the options. This application is, in our opinion, particularly interesting because it represents one of those cases in which the differential conditions are known a priori (given by theoretical arguments) and need to be taken into account in order to obtain valid estimates.\\
The raw data and the estimated smoothing surface (obtained through a soft constrained frequentist approach) are shown in figure~\ref{fig_BS_estimates}. These results have been obtained using tensor product B-splines of fourth order built on 25 equidistant internal knots for both the maturity and moneyness directions. Table~\ref{table_BS_results} shows the estimates of the implied volatility and related confidence/credibility intervals obtained both in frequentist and Bayesian settings.
The estimated implied volatilities in Table~\ref{table_BS_results} are in agreement with the median of the ones observed in the dataset.
The estimates obtained using our approach shows some desirable properties. First of all, we are able to extrapolate the prices associated with strikes and maturities not traded, and those estimates are consistent with the no arbitrage constraints (due to the adhesion to the Black and Scholes model). Furthermore, it is particularly appreciable in real data analysis to have confidence/credible bounds for prices associated to unobserved strikes and/or maturities. On the other hand, the PDE-P-spline approach allows here to obtain point and intervals estimates of the implied volatility consistent with the Black and Scholes equation ensuring a good performance in terms of data fitting. Frequentist and Bayesian approaches suggest compatible volatility parameters with narrow confidence/credibility intervals. These results are consistent with the ones obtained in our simulations. The smoothing surface given in Figure~\ref{fig_BS_estimates} is compatible with the observed cloud of prices.

\begin{table}[htbp]
  \centering
  \caption{Estimated implied volatility and confidence/credibility interval. The confidence interval is based on 1000 bootstrap replicates.}
    \begin{tabular}{cccc}
		\cline{2-4}
          & Estimates & 2.5\% & 97.5\% \\
		\hline \\
				Frequentist & 1.003E-01 & 9.940E-01 &  1.012E-01\\
				Bayesian & 1.014E-01 & 1.007E-01 & 1.020E-01\\
		\hline
		\end{tabular}
		\label{table_BS_results}
\end{table}

\section{Discussion}
\label{section:discussion}

In this paper, we propose PDE-based penalized tensor product B-spline smoothing approach in order to solve and estimate unknown parameters in partial differential equations. Our aim was to introduce frequentist and a Bayesian procedures to analyze data which dynamics is defined over more than one dimension. Both approaches exploit a tensor product B-spline approximation of the state function while the consistency of the final estimates to the PDE is ensured by PDE-based soft constraints. The compromise between the data fitting and consistency to the PDE is tuned by an adhesion parameter. As in \citet{xunetal2013}, we estimate the spline coefficients, the PDE parameters and the precision of measurement using the information implied in the observation. If differential conditions are not included in the model, our approach and the one by \citet{xunetal2013} are equivalent except that the PDE-adhesion parameter is selected through an EM-like algorithm \citep{schall1991}. For the Bayesian approach, we use the same PDE based penalty as the one we introduced in the frequentist approach. In particular, our Bayesian formulation takes into account the functional dependence of the normalizing constant of the spline coefficient prior on PDE parameters. Note that prior confidence in the PDE model is introduced by a specific choice for the prior distribution of the adhesion parameter.\\
We are able to improve the quality of the estimates by introducing differential conditions as extra constraints using either a L2 penalty or Lagrange multipliers. The introduction of these conditions arises sometimes naturally from theoretical arguments and cannot be omitted during the estimation process.\\
We demonstrate through simulations that our PDE-P-spline approaches show desirable properties and that the inclusion of the differential conditions has an influence on the estimation performances, reducing the variability of the PDE parameters and improving their coverages. We also compare our Bayesian proposals with the one suggested by \citet{xunetal2013}. We found that, within their Bayesian approach, the introduction of an additional finite difference penalty help to reduce the bias in estimating the PDE parameters. Indeed, we think that this additional penalty tends to mitigate the effect of the miss-specified constant of normalization in the prior of the spline coefficients. On the other hand, the same extra term tends to produce smoothing surface not adherent to the state function for small sample sizes. These undesirable features are not shared by our Bayesian proposal. Indeed, even for small sample sizes and/or large level of noise, the use of the proper prior for the spline coefficients ensures good fitting and estimating performances.\\
As real data application, we analyze the SP-500 call option prices discussed in \citet{carmona2004}. Despite the definition of the penalty term, the analyzed problem represents a real situation where the differential conditions play a role in the estimation process ensuring that the smoothing surface is consistent with the no-arbitrage requirements.
We use the unrealistic Black and Scholes model to define a suitable PDE-based penalty even if, paying a price in terms of computational complexity and interpretability of the final results, it is possible to consider more sophisticated and realistic pricing models.\\
Our further research will focus on possible extension of the PDE-P-spline approach.
The volatility in the Black and Scholes model is supposed to be constant. This assumption is known as unrealistic. The first solution in order to deal with the time-depending volatility could be to consider a more realistic differential pricing model (e.g. the Heston model \citep{heston1993}). In our opinion, the same aim can be achieved by allowing the implied volatility parameter in the Black and Scholes equation to vary nonparametrically with the time (e.g. using P-spline representation).\\
In this paper, we deal with linear partial differential equations. Moving to nonlinear partial differential equation would lead to additional challenges. First of all, the PDE-penalty term would not be a second order homogeneous polynomial in the spline coefficients. In the frequentist approach, this would require the use of the implicit function theorem to estimate the PDE parameters \citep{ramsayetal2007}. On the other hand, in a Bayesian setting, the constant of normalization for the prior distribution for the spline coefficients will not have an explicit form anymore.


\section*{Acknowledgment}
Support from the IAP Research Network P7/06 of the Belgian State (Belgian Science Policy) is gratefully acknowledged.


\bibliographystyle{plainnat}
\bibliography{biblio}


\appendix

\section{Logarithm of the marginalized posterior distribution}
\label{appendix:log_marg_post_dist}

The joint posterior distribution given in Eq.~\eqref{eq:joint_post_dist} is marginalized with respect to the spline coefficients. The log of the marginalized posterior distribution can be shown to be:
\begin{eqnarray*}
\log\left(p\left(\boldsymbol{\theta}, \gamma, \tau | \boldsymbol{\zeta}\right)\right) & = & \displaystyle\frac{N}{2} \log\left(\tau\right) - \displaystyle\frac{\tau}{2} \boldsymbol{\zeta}^{\top}\boldsymbol{\zeta}\\
                                                                                      & + & \displaystyle\frac{1}{2} \log\left(\mbox{det}\left(V_{1}\right)\right) - \displaystyle\frac{1}{2}\boldsymbol{v_{1}}^{\top}\boldsymbol{V_{1}}^{-1}\boldsymbol{v_{1}} \\
                                                                                      & - & \displaystyle\frac{1}{2} \log\left(\mbox{det}\left(V_{2}\right)\right) + \displaystyle\frac{1}{2}\boldsymbol{v_{2}}^{\top}\boldsymbol{V_{2}}^{-1}\boldsymbol{v_{2}} \\
																																											& + & \left(a_{\tau} - 1\right) \log\left(\tau\right) - b_{\tau}\tau \\
																																											& + & \left(a_{\gamma} - 1\right) \log\left(\gamma\right) - b_{\gamma} \gamma \\
																																											& + & \log\left(p\left(\boldsymbol{\theta}\right)\right).
\end{eqnarray*}

\section{Posterior distributions including the DE conditions by least squares}
\label{appendix:post_dist_least_squares}

The log joint posterior distribution can be shown to be:
\begin{eqnarray*}
\log\left(p\left(\boldsymbol{c}, \boldsymbol{\theta}, \gamma, \kappa, \tau | \boldsymbol{\zeta}\right)\right) & = & \displaystyle\frac{N}{2} \log\left(\tau\right) - \displaystyle\frac{\tau}{2} \left\|\boldsymbol{\zeta} - \boldsymbol{\mathcal{B}c} \right\| ^ {2} \\
                                                                                                      & + & \displaystyle\frac{1}{2} \log\left(\mbox{det}\left(V_{1}\right)\right) - \displaystyle\frac{1}{2}\left\{\boldsymbol{c}^{\top}\boldsymbol{V_{1}c} - 2 \boldsymbol{c}^{\top}\boldsymbol{v_{1}} + \boldsymbol{v_{1}}^{\top}\boldsymbol{V_{1}}^{-1}\boldsymbol{v_{1}}\right\} \\
																																																			& + & \left(a_{\tau} - 1\right) \log\left(\tau\right) - b_{\tau}\tau \\
																																																			& + & \left(a_{\gamma} - 1\right) \log\left(\gamma\right) - b_{\gamma} \gamma \\
																																																			& + & \left(a_{\kappa} - 1\right) \log\left(\kappa\right) - b_{\kappa} \kappa \\
																																																			& + & \log\left(p\left(\boldsymbol{\theta}\right)\right).
\end{eqnarray*}
From this joint posterior distribution, only two conditional posterior distribution can be identified. The conditional posterior distribution for the spline coefficients is a multivariate normal distribution:
\begin{equation*}
\boldsymbol{c} | \boldsymbol{\theta}, \tau, \gamma, \boldsymbol{\zeta} \sim \mathcal{N}_{M^{p}}\left(\boldsymbol{V_{2}}^{-1}\boldsymbol{v_{2}}; \boldsymbol{V_{2}}^{-1}\right).
\end{equation*}
where $\boldsymbol{V_{2}} = \tau \boldsymbol{\mathcal{B}}^{\top}\boldsymbol{\mathcal{B}} + \boldsymbol{V_{1}}$ and $\boldsymbol{v_{2}} = \tau \boldsymbol{\mathcal{B}}^{\top} \boldsymbol{\zeta} + \boldsymbol{v_{1}}$. For the precision of measurement, the conditional posterior distribution is a gamma distribution:
\begin{equation*}
\tau | \boldsymbol{c}, \boldsymbol{\zeta} \sim \mathcal{G}\left(\frac{N}{2} + a_{\tau}; \frac{\left\|\boldsymbol{\zeta} - \boldsymbol{\mathcal{B}c} \right\| ^ {2} }{2} + b_{\tau}\right).
\end{equation*}
The joint posterior distribution can be marginalized with respect to the spline coefficients. The log of the marginalized posterior distribution can be shown to be:
\begin{eqnarray*}
\log\left(p\left(\boldsymbol{\theta}, \gamma, \tau | \boldsymbol{\zeta}\right)\right) & = & \displaystyle\frac{N}{2} \log\left(\tau\right) - \displaystyle\frac{\tau}{2} \boldsymbol{\zeta}^{\top}\boldsymbol{\zeta}\\
                                                                                      & + & \displaystyle\frac{1}{2} \log\left(\mbox{det}\left(V_{1}\right)\right) - \displaystyle\frac{1}{2}\boldsymbol{v_{1}}^{\top}\boldsymbol{V_{1}}^{-1}\boldsymbol{v_{1}} \\
                                                                                      & - & \displaystyle\frac{1}{2} \log\left(\mbox{det}\left(V_{2}\right)\right) + \displaystyle\frac{1}{2}\boldsymbol{v_{2}}^{\top}\boldsymbol{V_{2}}^{-1}\boldsymbol{v_{2}} \\
																																											& + & \left(a_{\tau} - 1\right) \log\left(\tau\right) - b_{\tau}\tau \\
																																											& + & \left(a_{\gamma} - 1\right) \log\left(\gamma\right) - b_{\gamma} \gamma \\
																																											& + & \left(a_{\kappa} - 1\right) \log\left(\kappa\right) - b_{\kappa} \kappa \\
																																											& + & \log\left(p\left(\boldsymbol{\theta}\right)\right).
\end{eqnarray*}

\section{Elementary penalty elements}
\label{appendix:elementary_penalty_elements}
The penalty term, dealing with linear PDEs, can be seen as a homogeneous polynomial of second degree in the spline coefficients. The computations needed to construct the penalty term are:  
\begin{eqnarray*}
PEN(\mathbf{c}|\boldsymbol{\theta}) &=& \boldsymbol{c}^{\top} \boldsymbol{R}\left(\boldsymbol{\theta}\right) \boldsymbol{c} + 2 \boldsymbol{c}^{\top} \boldsymbol{r}\left(\boldsymbol{\theta}\right) + l\left(\boldsymbol{\theta}\right).
\end{eqnarray*}
The $\boldsymbol{R}$ matrix is given by the sum of a series of elementary penalty matrices:
\begin{eqnarray*}
\mathbf{P}_{i_{p},...,i_{1}}^{j_{p},...,j_{1}} &=& \int \left(\mathbf{B}_{p}^{(i_{p})} (x_{p}) \otimes \cdots \otimes \mathbf{B}_{1}^{(i_{1})}(x_{1})\right)  \left(\mathbf{B}_{p}^{(j_{p})} (x_{p}) \otimes \cdots \otimes \mathbf{B}_{1}^{(j_{1})}(x_{1})\right)^{\top}dx_{p} \ldots dx_{1},\\
                                               &=& \int \mathbf{B}_{p}^{(i_{p})} (x_{p}) \left(\mathbf{B}_{p}^{(j_{p})} (x_{p})  \right)^{\top}dx_{p} \otimes \cdots \otimes \int \mathbf{B}_{1}^{(i_{1})} (x_{1}) (\mathbf{B}_{1}^{(j_{1})} (x_{1}) )^{\top} dx_{1},\\
																							 &=& \mathcal{S}_{p}^{(i_p, j_p)} \otimes \cdots \otimes \mathcal{S}_{1}^{(i_1, j_1)},
\end{eqnarray*}
where the matrix $\mathcal{S}_{h}^{(i_h, j_h)}$ can be approximated using a trapezoidal rule. The $\boldsymbol{r}$ vector can be constructed as the sum of elementary penalty vectors:
\begin{eqnarray*}
\mathbf{p}_{i_{p},...,i_{1}} &=& \int \mathbf{B}_{p}^{(i_{p})} (x_{p}) \otimes \cdots \otimes \mathbf{B}_{1}^{(i_{1})}(x_{1})  dx_{p} \ldots dx_{1},\\
                                              &=& \int \mathbf{B}_{p}^{(i_{p})} (x_{p}) dx_{p} \otimes \cdots \otimes \int \mathbf{B}_{1}^{(i_{1})} (x_{1}) dx_{1},\\
																							 &=& \mathbf{s}_{p}^{(i_p)} \otimes \cdots \otimes \mathbf{s}_{1}^{(i_1)},																							
\end{eqnarray*}
as before, a trapezoidal rule can be applied to approximate the vector $\mathbf{s}_{h}^{(i_h)}$.
\end{document}